\documentclass[twocolumn,tighten,times]{aastex701}
\usepackage{amsmath}
\usepackage{amssymb}

\shorttitle{IXPE on LS I 61 303}
\shortauthors{Kaaret et al.}

\begin{document}

\title{X-Ray Polarization from the Gamma-Ray Binary LS I +61$^\circ$ 303}


\author[orcid=0000-0002-3638-0637,gname=Philip,sname=Kaaret]{Philip Kaaret}
\affiliation{NASA Marshall Space Flight Center, Huntsville, AL 35812, USA}
\email[show]{philip.kaaret@nasa.gov}

\author[0000-0003-4872-8159]{Sudip Chakraborty}
\affiliation{Science and Technology Institute, Universities Space and Research Association, Huntsville, AL 35805, USA}
\email{schakraborty2@usra.edu}

\author[0000-0003-4872-8159]{Daniel Golonka}
\affiliation{Science and Technology Institute, Universities Space and Research Association, Huntsville, AL 35805, USA}
\email{daniel.golonka@nasa.gov}

\author[0000-0002-7150-9061]{Oliver J. Roberts}
\affiliation{Science and Technology Institute, Universities Space and Research Association, Huntsville, AL 35805, USA}
\affiliation{School of Physics, University Road, University of Galway, Galway, Ireland, H91 TK33}
\email{oliver.roberts@universityofgalway.ie}

\author[0000-0001-9200-4006]{Ioannis Liodakis}
\affiliation{Institute of Astrophysics, Foundation for Research and Technology-Hellas, GR-70013 Heraklion, Greece}
\email{liodakis@ia.forth.gr}

\author[0000-0002-0642-1135]{Andrea Gnarini}
\affiliation{NASA Marshall Space Flight Center, Huntsville, AL 35812, USA}
\email{andrea.gnarini@uniroma3.it}

\author[0000-0003-4420-2838]{Steven R. Ehlert}
\affiliation{NASA Marshall Space Flight Center, Huntsville, AL 35812, USA}
\email{steven.r.ehlert@nasa.gov}

\author[0000-0001-7532-8359]{Joel B. Coley}
\affiliation{Department of Physics and Astronomy, Howard University, Washington, DC 20059, USA} 
\affiliation{CRESST and NASA Goddard Space Flight Center, Astrophysics Science Division, 8800 Greenbelt Road, Greenbelt, MD}
\email{joel.coley@howard.edu}

\begin{abstract}

The gamma-ray emitting binary stellar system LS~I~$+61^{\circ}$~303 was observed with the Imaging X-ray Polarimetry Explorer (IXPE) on two successive orbits over orbital phases of 0.74 to 1.05. Polarization is detected at a significance of $4.2\sigma$ with an average polarization degree of $13.1\% \pm 3.0\%$ in the 2-8~keV band after background subtraction. This is the second detection of polarization of the X-ray synchrotron emission from a gamma-ray binary and, again, suggests that the magnetic field in the particle acceleration region has a significant ordered component. The orbital motion on the sky of  LS~I~+61$^{\circ}$~303 is not well determined, which leads to ambiguity in interpretation of the X-ray electric vector polarization angle (EVPA) measurement. Use of orbital elements determined via radial velocity measurements combined with radio imaging of variable nebular emission, suggests an offset between the X-ray EVPA and the compact object-massive star axis on the order of $\sim 30^{\circ}$. Such an offset could be produced by Coriolis forces due to binary motion. Use of two different sets orbital elements determined via optical polarimetry suggest either no offset or a perpendicular orientation, but require an unexpectedly high inclination. Use of orbital elements derived from modeling of the keV/TeV light curves suggest good alignment between the X-ray EVPA and the compact object-massive star axis. Such alignment was found for the gamma-ray binary PSR B1259-63. If the same physical situation holds for LS~I~+61$^{\circ}$~303, that would favor the orbital elements derived from the keV/TeV light curves.

\end{abstract}


\section{Introduction}
\label{sec:intro}

Gamma-ray binaries are stellar binary systems emitting radiation at energies above 100~MeV and containing a compact object and a massive star \citep{Dubus2013}. The high energy radiation indicates the presence of highly relativistic particles. Radio pulsations on timescales of ms have been found in three gamma-ray binaries confirming that the compact object is a rapidly spinning neutron star: PSR B1259-63 \citep{Johnston1994}, PSR J2032$+$4127 \citep{Camilo2009}, and LS~I~$+61^{\circ}$~303 \citep{Weng2022}. 

The intrabinary shock formed where the relativistic outflow from a rapidly-rotating neutron star interacts with the wind from a massive star is a natural site for acceleration of particles to highly relativistic energies. The X-ray emission is then synchrotron radiation from these particles moving in the magnetic fields in or near the acceleration region \citep{Tavani1997}. X-ray polarization can be a useful tool to probe the magnetic field configuration near the acceleration region. 

The first such observations were made of the gamma-ray binary PSR~B1259–63 using the Imaging X-Ray Polarimetry Explorer (IXPE; \citealt{Weisskopf2022}) during an X-ray bright phase shortly after its periastron passage on 2024 June 30~\citep{Kaaret2024}. X-ray polarization was detected with a polarization degree of 8.3 $\pm$ 1.5\% at a significance of $5.3 \sigma$. The X-ray polarization angle was found to be aligned with the shock cone axis at the time of the observation, indicating the predominant component of the magnetic field in the acceleration region was oriented perpendicular to the shock cone axis. Measurements of other TeV binaries are needed to determine whether this magnetic field geometry is common for these kind of sources. The results can be used to advance our understanding of relativistic shock acceleration. 

Here, we present the results of IXPE observations of the gamma-ray binary LS~I~$+61^{\circ}$~303, hereafter LS61. We discuss some details of the binary in section~\ref{sec:ls61}, the IXPE observations in section~\ref{sec:obs}, and the results in section~\ref{sec:results}. We conclude, in section~\ref{sec:discussion}, with some discussion of the implications of the results.

\section{TeV Binary LS I +61 303}
\label{sec:ls61}

LS61 is a gamma-ray binary located at a distance of 2.65 $\pm$ 0.09~kpc \citep{Lindegren2021} and comprises a compact object in an eccentric orbit around a rapidly-rotating B0Ve star with an orbital period of $26.4960 \pm 0.0028$~d and a superorbital period of $1667 \pm 8$~d \citep{Gregory2002}. The best determinations of the periods are from long-term radio monitoring. The zero of the orbital phase of the compact object is set for historical reasons at MJD 43366.275 \citep{Gregory2002}.


The gamma-ray source 2CG~135+01 was discovered by the COS B satellite in a survey of the Galactic plane at energies above 100~MeV \citep{Hermsen1977}. In 2006, the MAGIC telescope provided greatly improved localization of the gamma-ray emission, confirming the long suspected identification with the optical and radio source LS~I~$+61^{\circ}$~303 \citep{Albert2006}. MAGIC also detected variable TeV gamma-ray emission modulated by the orbital period, which was confirmed by VERITAS observations \citep{Acciari2008}, with both instruments detecting enhanced TeV emission over orbital phases of 0.6 to 0.8. LS61 has also been observed to produce bright TeV flares with rise and fall times of less than a day \citep{Archambault2016}. Flares in the X-ray band are even faster with doubling times as fast as 2~s \citep{Smith2009}. The X-ray emission also shows a strong correlation between the flux and photon index, with the spectrum becoming harder at higher fluxes \citep{Smith2009,Li2011}. Recently, ultra-high energy emission ($>$100~TeV) was reported by the Large High Altitude Air Shower Observatory (LHASSO) with significances of 9.2$\sigma$ in the Water Cherenkov Detector Array (WCDA; 1.4–30.5 TeV) and 6.2$\sigma$ in the Kilometer Square Array (KM2A; 25–267 Tev) \citep{Cao2026}. The short time scales of X-ray \citep{Smith2009,Torres2012} and TeV flaring \citep{Archambault2016} suggests that particle acceleration can occur on short time scales, while the detection at ultra-high energies requires acceleration to ultra-high energies. 

Very Long Baseline Interferometry (VLBI) radio imaging reveals that the emission  from LS61 is extended on scales of milliarcseconds with time variable position and morphology. The position of the radio emission peak moves in a elliptical pattern as a function of orbital phase \citep{Dhawan2006,Wu2018}. However, the ellipse is much larger (semimajor axis $\approx 1.6$~mas) than the optically-measured orbit size (semimajor axis $\approx 0.2$~mas).

The nature of the compact object, either neutron star or black hole, in LS61 has long been debated. While the non-thermal emission from LS 61 was postulated soon after discovery to possibly come from a millisecond pulsar \citep{MaraschiTreves1981}, dedicated searches using the entire archive of Rossi X-ray Timing Explorer Proportional Counter Array (PCA) data revealed no pulsations \citep{Miralles2023}. \citet{Torres2012} reported a singular, short (0.24~s) X-ray burst they described as `magnetar-like' detected by the Swift Burst Alert Telescope and likely originating from LS61 which would suggest the compact object is a highly magnetized neutron star. Recent observations with the Five-hundred-meter Aperture Spherical radio Telescope (FAST) suggest that the compact object is a rotationally-powered NS with a period of $269.15508 \pm 0.00016$~ms \citep{Weng2022}. The FAST results are not fully conclusive because the field of view is too large to definitively rule out an interloper and no Doppler shift was detected as might be expected for a pulsar in a binary system \citep{Jaron2024}. However, the high significance of the pulsed signal ($> 20 \sigma$), the lack of accretion signatures such as disk emission, and the identification of pulsars in other gamma-ray binaries with larger orbital separation lead us to conclude the the compact object in LS61 is most likely a neutron star.

In this case, the particle acceleration likely occurs at the intrabinary shock formed where the relativistic wind from the neutron star interacts with the equatorial wind from the Be-star 
\citep{MaraschiTreves1981}. Also, the extended radio emission is likely synchrotron emission from particles accelerated in the same shock \citep{Dhawan2006}.


\subsection{Orbital elements}
\label{sec:orbit}

Knowledge of the position angle on the sky of the compact object relative to the binary system center of mass is needed to interpret the X-ray polarization angle. The orbit of PSR~B1259--63 was measured via highly accurate radio timing and astrometric observations of the pulsed emission which is detected over most of the orbit, except close to periastron \citep{MillerJones2018}. This provided direct measurement of the position of the pulsar itself. In contrast, pulsed emission from LS61 has been detected only in a single, 3-hr observation. Thus, no direct measurement of the orbital motion of the pulsar on the sky is available. Orbital information on LS61 is available from several different techniques: radial velocity measurements, optical polarimetry, and modeling of the high energy emission. 

The most common are radial velocity measurements \citep{Hutchings1981,Casares2005,Grundstrom2007,Aragona2009}. These provide the mass function ($f(M)$), eccentricity ($e$), and phase ($\phi_P$) and argument ($\omega$) of periastron. However, the optical line profiles are not purely from the atmosphere of the Be star, as assumed for radial velocity measurement, but may have contributions from the disk or gaseous shell surrounding the star. This requires careful selection of (relatively) uncontaminated absorption line features. Even so, the radial velocity measurements tend to exhibit large scatter around the best-fitted curve. Also, different solutions are found depending on the data set used, e.g.\ the eccentricity varies from $0.34 \pm 0.08$ to $0.72 \pm 0.15$ \citep{Grundstrom2007,Casares2005}.

Radial velocity measurements do not provide the orbital inclination ($i$) and longitude of ascending node ($\Omega$) which is associated with the orientation of the orbital major axis on the sky. The mass function, $f = M_\odot = (M_2 \sin i)^3/(M_1 + M_2)^2$ can be used to place some constraints on $i$. We use the mass function of $f = 0.0124 \pm 0.0022$ from \citet{Aragona2009} and assume the compact object is a neutron star with $M_2 = 1.35 M_\odot$ \citep{Ozel2016}. For a nominal B0Ve star mass of $M_1 = 12.5 M_\odot$, the inclination is then $i = 75^\circ$. We use this value below for orbital solutions that do not determine the inclination. The inclination increases with increasing $M_1$ and reaches $i = 90^\circ$ for Be star mass $M_1 \lesssim  13 M_\odot$. For Be star masses at the low end of the allowed range, $M_1 = 10 M_\odot$ \citep{Casares2005}, then $i = 40^\circ$. \citet{Hutchings1981} reported detection of `shell features' that are seen when the Be star is viewed through its disk and suggest that $\sin i$ is close to 1 under the assumption that Be disk is coplanar with the orbit. However, \citet{Casares2005} report a lack of clear shell lines and suggest that $i \lesssim 60^\circ$. The absence of X-ray eclipses would favor lower inclinations, $i \lesssim 60^\circ$, if the X-ray emission is confined to near the compact object.

To derive $\Omega$, we assume that ellipse traced by the radio emission peak is aligned with the binary orbit. \citet{Wu2018} derive a position angle (PA) of $\theta_{\text{obs}} \approx -43.6^\circ \pm 1.6^\circ$ East of North for the radio peak ellipse. Converting this to the standard $0^\circ-180^\circ$ convention for the line of nodes yields a position angle for the major axis of $\theta = 136.4^\circ$. The apparent orientation of the projected semi-major axis on the sky ($\theta$) is related to $\Omega$, $i$, and $\omega$ through the 3D projection of the periastron vector onto the sky plane. In Appendix~\ref{sec:ascending}, we show that the longitude of the ascending node is $\Omega = \theta - \arctan(\tan \omega \cos i)$. Thus, given values $i$ and $\omega$, we can calculate the $\Omega$ from the \citet{Wu2018} estimation of $\theta$.


\citet{Kravtsov2020} use a fundamentally different technique, high-precision optical photo-polarimetry, to derive the orbital elements of LS61. Because polarimetry provides geometric information, the measurements do accurately constrain the orientation of the orbit on the sky and, thus, use of the VLBI imaging is not needed. Optical polarimetry is nominally sensitive to the orbital inclination. However, the inclination results are biased towards high values in the presence of instrumental or intrinsic noise as described further below.

The time-dependent high energy emission observed from gamma-ray binaries is strongly influenced by the orbital geometry. \citet{Chen2022} developed a model for the emission (X-ray synchtron and gamma-ray inverse Compton) produced by particles accelerated at the intrabinary shock in binaries containing a pulsar orbiting a Be star. They used the model to estimate the orbital elements for LS61 via fits to the observed X-ray and gamma-ray light curves. The light curves do not include geometric information. Thus, we need to add an assumed $i$ and calculate $\Omega$ from the VLBI imaging, as for the radial velocity orbital solutions.

\section{IXPE Observations and Reduction}
\label{sec:obs}

IXPE observed LS61 in three segments from 14 Feb 2026 (MJD 61085.250) to 19 Mar 2026 (MJD 61118.217). IXPE has three X-ray telescopes, each containing a Mirror Module Assembly \citep[][MMA]{Ramsey2022} and a detector unit (DU) housing a gas pixel detector (GPD) sensitive to linear X-ray polarization  \citep{Soffitta2021,Baldini2021}. We analyze data from all three DUs. We removed background flares by filtering on 2-8~keV band counting rates extracted from a background region with inner radius of 2.5$\arcmin$ and outer radius of 5.0$\arcmin$, excising intervals when summed count rate in the three DUs in a 240~s time bin exceeded the mean rate by more than 3 times the standard error on the rate measurement. After filtering, the total exposure was 713.9, 713.7, and 714.0~ks for DU1, DU2, and DU3, respectively. 

We extracted source counts from a circular region with a radius of $60\arcsec$. The source region was centered visually using the IXPE images; the best centroid was 7.5$\arcsec$ from the nominal source location. We extracted background counts from a concentric annulus with radii of $150\arcsec$ and $300\arcsec$ \citep{DiMarco2023}. The source spectrum is above the background spectrum, scaled for the region sizes, in the full IXPE 2-8~keV band. Hence, we use that energy band for subsequent analysis. The data were processed with CALDB version 20260331 and HEASoft version 6.36 was used in the analysis.

\begin{figure}[tb]
\centerline{\includegraphics[width=3.0in]{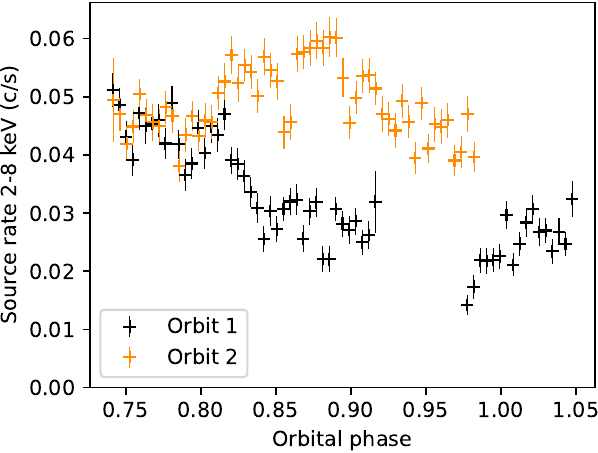}}
\caption{IXPE count rate in the 2-8~keV band versus orbital phase. Black points are observations during the first orbit, while orange points are during the second orbit.}
\label{fig:flux_phase}
\end{figure}

\section{Results}
\label{sec:results}

The IXPE observations cover portions of two orbits of LS61. Figure~\ref{fig:flux_phase} shows the IXPE count rate in the 2-8~keV band versus orbital phase. The time bins are $10^4$~s and the rate is summed over the 3 DUs. The phase coverage was planned to match times when the X-ray flux is brighter and TeV emission has historically been observed \citep{Smith2009}. The flux was notably higher during the second orbit.

\begin{figure}[tb]
\centerline{\includegraphics[width=3.0in]{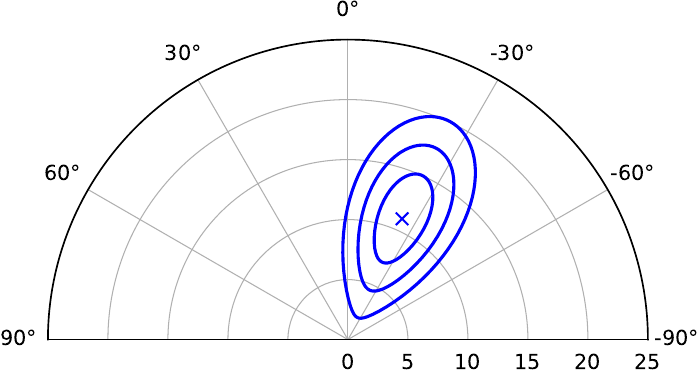}}
\caption{Contour plots of the polarization degree and angle without correction for background. The $\times$ marks the measured values. The contours show the 68.27\%, 95.45\%, and 99.73\% confidence intervals for $\chi^2$ on 2 degrees of freedom. Position angles are measured positive east of north.}
\label{fig:protractor}
\end{figure}

For our initial polarization analysis, we performed a model-independent calculation in the 2-8~keV band using counts from the source region for the full observation. Each event was weighted based on how well the electron track could be used to correctly reconstruct the original photoelectron direction. No correction was made for background. Figure~\ref{fig:protractor} shows the average polarization of the source events. The plot was produced with the \texttt{ixpe\_protractor.py} script on the IXPE User Contributed software page, which uses the \texttt{ ixpepolarization} tool in HEASoft. The polarization results from the individual DUs are all consistent with the summed result within the 68\% confidence contour.

The observed polarization degree (PD) is $11.0\% \pm 2.6\%$. All uncertainties are reported at 68.27\% confidence. The detection is significant at the $4.2 \sigma$ confidence level and indicates a secure detection of X-ray polarization.  The electric vector position angle (EVPA) is $-24\fdg1 \pm  6\fdg8$ and is measured anticlockwise from north in the equatorial coordinate system. We found no significant polarization for the background region. Correcting for background utilizing the additive nature of the Stokes parameters, we estimate the PD of the source as $13.1 \% \pm 3.0\%$ and the EVPA as $-23\fdg8 \pm 6\fdg6$. The EVPA is consistent with that found without background subtraction. The source region contained 33284 counts, of which we estimate 4258 are due to background. Due to the relatively low background fraction, we did not attempt to reduce the instrumental background using the rejection algorithm described in \citet{DiMarco2023}.

Motivated by the change in flux between the two orbits, we found the polarization for each orbit. We find no statistically significant difference. For the first orbit, $\rm PD = 11.2\% \pm 4.0\%$, and for the second orbit, $\rm PD = 11.1\% \pm 3.5\%$. The EVPA is consistent between the two orbits and with the observation average. We also divided the observation into two intervals by flux; there was no significant difference in PD.


We performed a spectropolarimetric analysis. We generated I, Q, and U spectra in the 2-8~keV band and response files for each DU using \texttt{ixpeproduct} in HEASoft. The same source and background regions described above were used. The Stokes $I$ spectra were binned to have 80~eV bins below 3~keV and  120~eV above. The Stokes $Q$ and $U$ spectra were binned with 400~eV bins. We modeled the spectrum using an absorbed power law with constant polarization multiplied by a constant allowed to vary between the three DUs. The column density for the \texttt{phabs} absorption model was fixed to $0.5 \times 10^{22}\,\rm cm^{-2}$ \citep{Chernyakova2017} using abundances from \citet{Anders1989} and cross-sections from \citet{Verner1996}. We obtained a good fit with $\chi^2/\rm DoF = 238.6/240$. We found $\rm PD = 11.4\% \pm 2.8\%$ and $\rm EVPA = -24\fdg1 \pm  7\fdg1$. Both are consistent with the background-subtracted model-independent results. We found a photon index of $\Gamma = 1.46 \pm 0.03$ within the range measured previously in the 2-10~keV band \citep{Smith2009}.

\begin{figure}[tb]
\centerline{\includegraphics[width=3.25in]{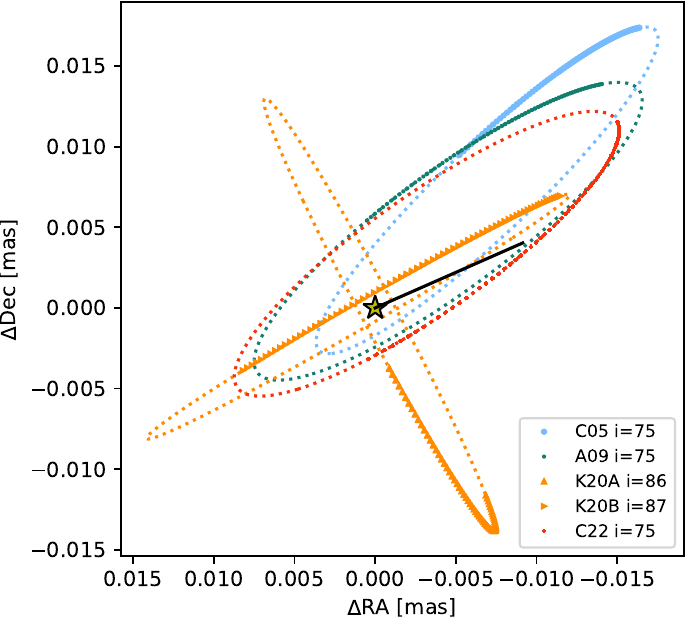}}
\caption{Orbital motion of LS61 on the sky. Shown is the orbital motion of LS61 for the orbital solution of Casares (C05, blue), Aragona (A09, green), and Chen (C22, orange) for $i = 60^{\circ}$ and the orbital solution of Kravtkov (K20, vermillion). The star shows the system barycenter. The black line segment emanating from the star indicates the X-ray EVPA. The points mark the times of the IXPE light curve segments shown in Fig.~\ref{fig:flux_phase}.}
\label{fig:orbit}
\end{figure}

\section{Discussion}
\label{sec:discussion}

Our IXPE observations have led to the detection of significant X-ray polarization ($4.2 \sigma$ confidence level) from LS61 with a background-subtracted $\rm PD = 13.1 \% \pm 3.0\%$. This is the second detection of X-ray polarization from a TeV binary. The PD is higher than that detected from PSR~B1259–63 during an X-ray bright phase following the periastron passage in 2024 June \citep{Kaaret2024}.

The polarization degree (PD) for synchrotron radiation is determined by the particle spectral index and the level of uniformity of the magnetic field. The maximum polarization for LS61 is $\sim 70\%$. The reduction in observed PD relative to the maximum is roughly equal to to the ratio of the energy in the uniform magnetic field component to that in the total field \citep{Burn1966}. The observed PD of 13\% suggests that $\sim 20\%$ of the magnetic field is in the uniform component. However, the observed PD can also be reduced if the shock-cone axis is not perpendicular to the line of sight \citep{Laing1980,Xingxing2021} or due to curvature of the shock in the acceleration region. Thus, the fraction of $\sim 20\%$ of the field in the uniform component should be considered as a lower bound.

\begin{figure}[tb]
\centerline{\includegraphics[width=3.25in]{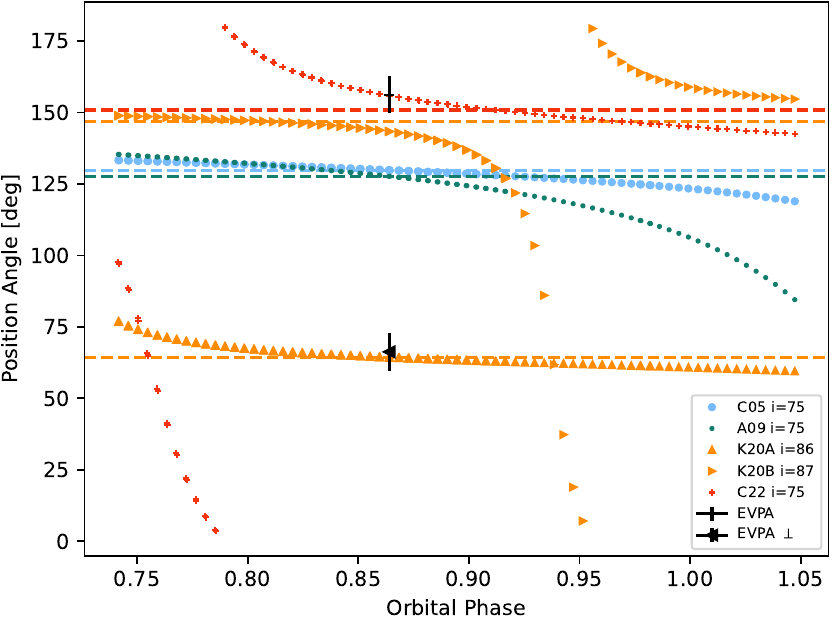}}
\caption{Pulsar position angle versus orbital phase for the IXPE light curve segments. The orbital solutions and inclinations are as in Fig.~\ref{fig:orbit}. The dashed horizontal lines indicate the median of each distribution. The points show the X-ray EVPA and rotated by $90^\circ$ (EVPA $\perp$) and are plotted at the median orbital phase of the IXPE light curve segments.}
\label{fig:paphase}
\end{figure}

The X-ray EVPA for synchrotron radiation is set by the magnetic field orientation in and near the acceleration region. This is set by the shock cone axis which thought to be is aligned with the pulsar orbital position angle on the sky at the time of the observation \citep{Bogovalov2008}. Figure~\ref{fig:orbit} shows the pulsar orbit for LS61 projected on the sky calculated with orbital parameters determined as described in section~\ref{sec:orbit}. We show orbits using the elements from \citet{Casares2005}, \citet{Aragona2009}, and \citet{Chen2022} for which we have taken $i = 75^\circ$ and using the two solutions from \citet{Kravtsov2020} with the fitted inclination indicated in the legend. In the figure, the dotted curves show the full orbit. The times of the IXPE light curve points shown in Fig.~\ref{fig:flux_phase} are plotted as points. The star shows the system center of mass with the line segment emanating from it indicating the X-ray EVPA.

The pulsar moves over a range of position angles during the IXPE observations. The position angle versus orbital phase for each light segment is shown in Figure~\ref{fig:paphase}. The measured X-ray EVPA of $180 -23\fdg8 \pm 6\fdg6 = 156\fdg2 \pm 6\fdg6$ is also shown. Remarkably, the EVPA is not aligned with the pulsar position angle during the observation for either the Casares or Aragona orbital elements. The minimum offset between EVPA and median is $26^\circ$ for the Casares elements with $i = 75^\circ$. The offset for the Aragona elements is $28^\circ$. These offsets are significantly larger than the uncertainty in the EVPA. Neither of these two position angle distributions overlap with the measured EVPA allowing for uncertainties. The offset increases for lower $i$.

In contrast, the offset between the X-ray EVPA and median pulsar position angle during the IXPE observations as calculated using the Kravtsov orbital elements with $\Omega$ free (lower part of Table~4, K20B in Figures~\ref{fig:orbit} and \ref{fig:paphase}) is only $10^\circ$. The EVPA and median pulsar position angle are consistent within the 90\% confidence interval. However, the best fitted inclination is high, $i = 87^\circ \pm 3^\circ$. Polarimetrically-derived inclinations are biased towards high values when noise is present \citep{Simmons1982}. The optical polarization signal from LS61 shows strong intrinsic stochastic variations that drive the fitted inclination to high values, but the $2\sigma$ confidence interval extends down to  $i = 20^\circ$ \citep{Kravtsov2020}. The pulsar position angle inferred from the Kravtsov orbital elements depends strongly on $i$, assuming that the other orbital elements remain unchanged if the inclination is varied (Kravtsov, private communication). For $i = 80^\circ$, the offset is as large as for the orbits based on radial velocity measurements.

In contrast, for the Kravtsov orbital elements with $\Omega$ constrained (upper part of Table~4, K20A in our figures), the median of the pulsar position angle is perpendicular to the X-ray EVPA within $2^\circ$. This orbit is largely perpendicular to the other orbits. Again, the offset increases for smaller $i$ and is as large as for the orbits based on radial velocity measurements for $i = 78^\circ$. 

Remarkably, the orbital elements of \citet{Chen2022} provide a good match to the X-ray EVPA when using $i = 75^\circ$ and the orbital orientation on the sky inferred from the radio imaging. The median pulsar position angle is $151^{\circ}$ which is well within the 68\% confidence interval for the X-ray EVPA. A good match is also obtained for $i = 60^\circ$, with a median pulsar position angle of $153.7^{\circ}$, only $2.5^\circ$ from the X-ray EVPA. A large offset of $99^\circ$ is obtained for $i = 45^\circ$.

For PSR~B1259–63, the X-ray polarization angle was found to be well aligned with the pulsar position angle during the IXPE observation \citep{Kaaret2024}. This was interpreted as implying that the axis of the shock cone, where the high energy particles producing the X-ray synchrotron emission are accelerated, was also aligned with the pulsar-primary axis and, therefore, that the predominant component of the magnetic field in the acceleration region is oriented perpendicular to the shock cone axis. If the same physical situation found for applies to LS61, then the IXPE results would favor the Chen orbit elements. 

The Kravtsov orbital elements found with $\Omega$ free (K20B) could also provide a good match, however the high orbital inclination appears to be an artifact, thus the match may be spurious. The Kravtsov orbital elements found with $\Omega$ constrained lead to a pulsar position angle nearly perpendicular to the X-ray EVPA. This would imply that the magnetic field is parallel to the shock cone axis, orthogonal to the geometry found for PSR~B1259–63. Again, the match may be spurious due to the unexpectedly high inclination.

If the orbits derived from radial velocity measurements and radio imaging are correct, then there is a misalignment on the orbit of $\sim 30^\circ$ between the pulsar-primary position angle and the X-ray EVPA. This would suggest that the shock region where particles are accelerated to produce the observed X-ray synchrotron emission is misaligned with the pulsar-primary axis. \citet{Bosch2012} have performed relativistic hydrodynamical simulations in two dimensions of the stellar/pulsar-wind interaction accounting for orbital motion. They find that the Coriolis force induced by orbital motion deflects the shocked flows. This causes the shock region to become asymmetric with respect to the pulsar-primary axis. In this case, measurement of the X-ray EVPA would provide a probe into the dynamics of the stellar-pulsar wind interaction.

X-ray polarimetry can provide important, new information on shock acceleration and the pulsar-stellar wind interaction. The X-ray polarization degree observed from LS61 suggests there is a substantial well-ordered component of the magnetic field within or near the shock acceleration region. The observed X-ray polarization angle is related to the pulsar motion projected on the sky  and may provide an independent way to break degeneracies and settle debates regarding the true orbit of LS61. If the true orbit can be determined independently, then the X-ray position angle would provide a probe of the intrabinary shock formation.

\appendix
\section{Longitude of ascending node}
\label{sec:ascending}

The position angle ($\theta$) of the projected major axis is the sum of the longitude of ascending node ($\Omega$) and the apparent, projected argument of periastron ($\alpha$), $ \theta = \Omega + \alpha $. To derive the $\alpha$, we consider the transformation from the physical orbital plane to the observer's sky plane. We define a coordinate system $(x', y', z')$ in the orbital frame where the $x'$-axis lies along the Line of Nodes and the orbit lies in the $(x', y')$ plane. The position vector of periastron at a distance $r$ is

\begin{equation}
\vec{v}_{\text{orb}} = \begin{bmatrix} r \cos \omega \\ r \sin \omega \\ 0 \end{bmatrix}
\end{equation}

\noindent where $\omega$ is the argument of periastron. To project this onto the sky plane $(x, y)$, we rotate the system by the inclination $i$ around the Line of Nodes ($x'$-axis),

\begin{equation}
\begin{bmatrix} x \\ y \\ z \end{bmatrix} = \begin{bmatrix} 1 & 0 & 0 \\ 0 & \cos i & -\sin i \\ 0 & \sin i & \cos i \end{bmatrix} \begin{bmatrix} r \cos \omega \\ r \sin \omega \\ 0 \end{bmatrix} = \begin{bmatrix} r \cos \omega \\ r \sin \omega \cos i \\ r \sin \omega \sin i \end{bmatrix}
\end{equation}

The observer on Earth perceives the projection $(x, y)$. The apparent angle $\alpha$ relative to the Line of Nodes ($x$-axis) is given by
\begin{equation}
\tan \alpha = \frac{y}{x} = \frac{r \sin \omega \cos i}{r \cos \omega} = \tan \omega \cos i
\end{equation}
Therefore,
\begin{equation}
\alpha = \arctan(\tan \omega \cos i)
\end{equation}

Solving for the longitude of the ascending node, we find

\begin{equation}
\Omega = \theta - \arctan(\tan \omega \cos i).
\end{equation}

\begin{acknowledgments}

We thank Vadim Kravtsov for discussion of his optical polarimetry results. OJR acknowledges funding through NASA and Research Ireland Pathway Funding under contract 24/PATH-S/12742(T). AG was supported by an appointment to the NASA Postdoctoral Program at the Marshall Space Flight Center (MSFC), administered by Oak Ridge Associated Universities under contract with NASA. JBC acknowledges support by NASA under award number 80GSFC21M0006. IL was funded by the European Union ERC-2022-STG - BOOTES - 101076343. Views and opinions expressed are however those of the author(s) only and do not necessarily reflect those of the European Union or the European Research Council Executive Agency. Neither the European Union nor the granting authority can be held responsible for them.

The Imaging X-ray Polarimetry Explorer (IXPE) is a joint US and Italian mission.  The US contribution is supported by the National Aeronautics and Space Administration (NASA) and led and managed by its Marshall Space Flight Center (MSFC), with industry partner Ball Aerospace (now, BAE Systems). The Italian contribution is supported by the Italian Space Agency (Agenzia Spaziale Italiana, ASI) through contract ASI-OHBI-2022-13-I.0, agreements ASI-INAF-2022-19-HH.0 and ASI-INFN-2017.13-H0, and its Space Science Data Center (SSDC) with agreements ASI-INAF-2022-14-HH.0 and ASI-INFN 2021-43-HH.0, and by the Istituto Nazionale di Astrofisica (INAF) and the Istituto Nazionale di Fisica Nucleare (INFN) in Italy. This research used data products provided by the IXPE Team (MSFC, SSDC, INAF, and INFN) and distributed with additional software tools by the High-Energy Astrophysics Science Archive Research Center (HEASARC), at NASA Goddard Space Flight Center (GSFC). 

\end{acknowledgments}





\facilities{IXPE}

\software{
\textsc{HEAsoft} \citep{HEAsoft},
\textsc{Xspec} \citep{xspec},
\textsc{ds9} \citep{ds9},
\textsc{AstroPy} \citep{astropy2013,astropy2018,astropy2022}
}

\bibliography{tevbinary}

@ARTICLE{Chen2022,
       author = {{Chen}, A.~M. and {Takata}, J.},
        title = "{Modelling the correlated keV/TeV light curves of Be/gamma-ray binaries}",
      journal = {\aap},
         year = 2022,
        month = feb,
       volume = {658},
          eid = {A153},
        pages = {A153},
          doi = {10.1051/0004-6361/202142258},
archivePrefix = {arXiv},
       eprint = {2112.00345},
 primaryClass = {astro-ph.HE},
       adsurl = {https://ui.adsabs.harvard.edu/abs/2022A&A...658A.153C}
}

@ARTICLE{Verner1996,
       author = {{Verner}, D.~A. and {Ferland}, G.~J. and {Korista}, K.~T. and {Yakovlev}, D.~G.},
        title = "{Atomic Data for Astrophysics. II. New Analytic Fits for Photoionization Cross Sections of Atoms and Ions}",
      journal = {\apj},
         year = 1996,
        month = jul,
       volume = {465},
        pages = {487},
          doi = {10.1086/177435},
archivePrefix = {arXiv},
       eprint = {astro-ph/9601009},
 primaryClass = {astro-ph},
       adsurl = {https://ui.adsabs.harvard.edu/abs/1996ApJ...465..487V}
}

@ARTICLE{Hutchings1981,
       author = {{Hutchings}, J.~B. and {Crampton}, D.},
        title = "{Spectroscopy of the unique degenerate binary star LS I +61 303.}",
      journal = {\pasp},
         year = 1981,
        month = aug,
       volume = {93},
        pages = {486-489},
          doi = {10.1086/130863},
       adsurl = {https://ui.adsabs.harvard.edu/abs/1981PASP...93..486H}
}

@ARTICLE{Kaaret2024,
       author = {{Kaaret}, Philip and {Roberts}, Oliver J. and {Ehlert}, Steven R. and {Swartz}, Douglas A. and {Weisskopf}, Martin C. and {Liodakis}, Ioannis and {Saade}, M. Lynne and {O'Dell}, Stephen L. and {Chen}, Chien-Ting},
        title = "{Magnetic Field Geometry of the Gamma-Ray Binary PSR B1259─63 Revealed via X-Ray Polarization}",
      journal = {\apjl},
         year = 2024,
        month = oct,
       volume = {974},
       number = {1},
          eid = {L1},
        pages = {L1},
          doi = {10.3847/2041-8213/ad7ba6},
archivePrefix = {arXiv},
       eprint = {2409.16116},
 primaryClass = {astro-ph.HE},
       adsurl = {https://ui.adsabs.harvard.edu/abs/2024ApJ...974L...1K}
}

@ARTICLE{Bosch2012,
       author = {{Bosch-Ramon}, V. and {Barkov}, M.~V. and {Khangulyan}, D. and {Perucho}, M.},
        title = "{Simulations of stellar/pulsar-wind interaction along one full orbit}",
      journal = {\aap},
         year = 2012,
        month = aug,
       volume = {544},
          eid = {A59},
        pages = {A59},
          doi = {10.1051/0004-6361/201219251},
archivePrefix = {arXiv},
       eprint = {1203.5528},
 primaryClass = {astro-ph.HE},
       adsurl = {https://ui.adsabs.harvard.edu/abs/2012A&A...544A..59B}
}

@ARTICLE{Anders1989,
       author = {{Anders}, E. and {Grevesse}, N.},
        title = "{Abundances of the elements: Meteoritic and solar}",
      journal = {\gca},
         year = 1989,
        month = jan,
       volume = {53},
       number = {1},
        pages = {197-214},
          doi = {10.1016/0016-7037(89)90286-X},
       adsurl = {https://ui.adsabs.harvard.edu/abs/1989GeCoA..53..197A}
}

@ARTICLE{Chernyakova2017,
       author = {{Chernyakova}, M. and {Babyk}, Iu. and {Malyshev}, D. and {Vovk}, Ie. and {Tsygankov}, S. and {Takahashi}, H. and {Fukazawa}, Ya.},
        title = "{Study of orbital and superorbital variability of LSI +61{\textdegree} 303 with X-ray data}",
      journal = {\mnras},
         year = 2017,
        month = sep,
       volume = {470},
       number = {2},
        pages = {1718-1728},
          doi = {10.1093/mnras/stx1335},
archivePrefix = {arXiv},
       eprint = {1705.09343},
 primaryClass = {astro-ph.HE},
       adsurl = {https://ui.adsabs.harvard.edu/abs/2017MNRAS.470.1718C}
}

@ARTICLE{Hermsen1977,
       author = {{Hermsen}, W. and {Swanenburg}, B.~N. and {Bignami}, G.~F. and {Boella}, G. and {Buccheri}, R. and {Scarsi}, L. and {Kanbach}, G. and {Mayer-Hasselwander}, H.~A. and {Masnou}, J.~L. and {Paul}, J.~A.},
        title = "{New high energy gamma-ray sources observed by COS B}",
      journal = {\nat},
         year = 1977,
        month = oct,
       volume = {269},
        pages = {494-495},
          doi = {10.1038/269494a0},
       adsurl = {https://ui.adsabs.harvard.edu/abs/1977Natur.269..494H}
}

@ARTICLE{Camilo2009,
       author = {{Camilo}, F. and {Ray}, P.~S. and {Ransom}, S.~M. and {Burgay}, M. and {Johnson}, T.~J. and {Kerr}, M. and {Gotthelf}, E.~V. and {Halpern}, J.~P. and {Reynolds}, J. and {Romani}, R.~W. and {Demorest}, P. and {Johnston}, S. and {van Straten}, W. and {Saz Parkinson}, P.~M. and {Ziegler}, M. and {Dormody}, M. and {Thompson}, D.~J. and {Smith}, D.~A. and {Harding}, A.~K. and {Abdo}, A.~A. and {Crawford}, F. and {Freire}, P.~C.~C. and {Keith}, M. and {Kramer}, M. and {Roberts}, M.~S.~E. and {Weltevrede}, P. and {Wood}, K.~S.},
        title = "{Radio Detection of LAT PSRs J1741-2054 and J2032+4127: No Longer Just Gamma-ray Pulsars}",
      journal = {\apj},
         year = 2009,
        month = nov,
       volume = {705},
       number = {1},
        pages = {1-13},
          doi = {10.1088/0004-637X/705/1/1},
archivePrefix = {arXiv},
       eprint = {0908.2626},
 primaryClass = {astro-ph.GA},
       adsurl = {https://ui.adsabs.harvard.edu/abs/2009ApJ...705....1C}
}

@ARTICLE{Weng2022,
       author = {{Weng}, Shan-Shan and {Qian}, Lei and {Wang}, Bo-Jun and {Torres}, D.~F. and {Papitto}, A. and {Jiang}, Peng and {Xu}, Renxin and {Li}, Jian and {Yan}, Jing-Zhi and {Liu}, Qing-Zhong and {Ge}, Ming-Yu and {Yuan}, Qi-Rong},
        title = "{Radio pulsations from a neutron star within the gamma-ray binary LS I +61{\textdegree} 303}",
      journal = {Nature Astronomy},
         year = 2022,
        month = mar,
       volume = {6},
        pages = {698-702},
          doi = {10.1038/s41550-022-01630-1},
archivePrefix = {arXiv},
       eprint = {2203.09423},
 primaryClass = {astro-ph.HE},
       adsurl = {https://ui.adsabs.harvard.edu/abs/2022NatAs...6..698W}
}

@INPROCEEDINGS{Dhawan2006,
       author = {{Dhawan}, Vivek and {Mioduszewski}, Amy and {Rupen}, M.},
        title = "{LS I +61 303 is a Be-Pulsar binary, not a Microquasar}",
    booktitle = {VI Microquasar Workshop: Microquasars and Beyond},
         year = 2006,
        month = jan,
          eid = {52.1},
        pages = {52.1},
          doi = {10.22323/1.033.0052},
       adsurl = {https://ui.adsabs.harvard.edu/abs/2006smqw.confE..52D}
}

@ARTICLE{Aragona2009,
       author = {{Aragona}, Christina and {McSwain}, M. Virginia and {Grundstrom}, Erika D. and {Marsh}, Amber N. and {Roettenbacher}, Rachael M. and {Hessler}, Katelyn M. and {Boyajian}, Tabetha S. and {Ray}, Paul S.},
        title = "{The Orbits of the {\ensuremath{\gamma}}-Ray Binaries LS I +61 303 and LS 5039}",
      journal = {\apj},
         year = 2009,
        month = jun,
       volume = {698},
       number = {1},
        pages = {514-518},
          doi = {10.1088/0004-637X/698/1/514},
archivePrefix = {arXiv},
       eprint = {0902.4015},
 primaryClass = {astro-ph.HE},
       adsurl = {https://ui.adsabs.harvard.edu/abs/2009ApJ...698..514A}
}

@ARTICLE{Kravtsov2020,
       author = {{Kravtsov}, Vadim and {Berdyugin}, Andrei V. and {Piirola}, Vilppu and {Kosenkov}, Ilia A. and {Tsygankov}, Sergey S. and {Chernyakova}, Maria and {Malyshev}, Denys and {Sakanoi}, Takeshi and {Kagitani}, Masato and {Berdyugina}, Svetlana V. and {Poutanen}, Juri},
        title = "{Orbital variability of the optical linear polarization of the {\ensuremath{\gamma}}-ray binary LS I +61{\textdegree} 303 and new constraints on the orbital parameters}",
      journal = {\aap},
         year = 2020,
        month = nov,
       volume = {643},
          eid = {A170},
        pages = {A170},
          doi = {10.1051/0004-6361/202038745},
archivePrefix = {arXiv},
       eprint = {2010.00999},
 primaryClass = {astro-ph.SR},
       adsurl = {https://ui.adsabs.harvard.edu/abs/2020A&A...643A.170K}
}

@ARTICLE{Simmons1982,
       author = {{Simmons}, J.~F.~L. and {Aspin}, C. and {Brown}, J.~C.},
        title = "{Bias of polarimetric estimators for binary star inclinations.}",
      journal = {\mnras},
         year = 1982,
        month = jan,
       volume = {198},
        pages = {45-57},
          doi = {10.1093/mnras/198.1.45},
       adsurl = {https://ui.adsabs.harvard.edu/abs/1982MNRAS.198...45S}
}

@ARTICLE{Casares2005,
       author = {{Casares}, J. and {Ribas}, I. and {Paredes}, J.~M. and {Mart{\'\i}}, J. and {Allende Prieto}, C.},
        title = "{Orbital parameters of the microquasar LS I +61 303}",
      journal = {\mnras},
         year = 2005,
        month = jul,
       volume = {360},
       number = {3},
        pages = {1105-1109},
          doi = {10.1111/j.1365-2966.2005.09106.x},
archivePrefix = {arXiv},
       eprint = {astro-ph/0504332},
 primaryClass = {astro-ph},
       adsurl = {https://ui.adsabs.harvard.edu/abs/2005MNRAS.360.1105C}
}

@ARTICLE{Gregory2002,
       author = {{Gregory}, P.~C.},
        title = "{Bayesian Analysis of Radio Observations of the Be X-Ray Binary LS I +61{\textdegree}303}",
      journal = {\apj},
         year = 2002,
        month = aug,
       volume = {575},
       number = {1},
        pages = {427-434},
          doi = {10.1086/341257},
       adsurl = {https://ui.adsabs.harvard.edu/abs/2002ApJ...575..427G}
}

@article{Cao2026,
  title = {First detection of ultrahigh energy emission from gamma-ray binary LS I +61${}^{\ensuremath{\circ}}$ 303},
  author = {Cao, Zhen and Aharonian, F. and Bai, Y. X. and Bao, Y. W. and Bastieri, D. and Bi, X. J. and Bi, Y. J. and Bian, W. and Bukevich, A. V. and Cai, C. M. and Cao, W. Y. and Cao, Zhe and Chang, J. and Chang, J. F. and Chen, A. M. and Chen, E. S. and Chen, G. H. and Chen, H. X. and Chen, Liang and Chen, Long and Chen, M. J. and Chen, M. L. and Chen, Q. H. and Chen, S. and Chen, S. H. and Chen, S. Z. and Chen, T. L. and Chen, X. B. and Chen, X. J. and Chen, Y. and Cheng, N. and Cheng, Y. D. and Chu, M. C. and Cui, M. Y. and Cui, S. W. and Cui, X. H. and Cui, Y. D. and Dai, B. Z. and Dai, H. L. and Dai, Z. G. and Danzengluobu and Diao, Y. X. and Dong, X. Q. and Duan, K. K. and Fan, J. H. and Fan, Y. Z. and Fang, J. and Fang, J. H. and Fang, K. and Feng, C. F. and Feng, H. and Feng, L. and Feng, S. H. and Feng, X. T. and Feng, Y. and Feng, Y. L. and Gabici, S. and Gao, B. and Gao, C. D. and Gao, Q. and Gao, W. and Gao, W. K. and Ge, M. M. and Ge, T. T. and Geng, L. S. and Giacinti, G. and Gong, G. H. and Gou, Q. B. and Gu, M. H. and Guo, F. L. and Guo, J. and Guo, X. L. and Guo, Y. Q. and Guo, Y. Y. and Han, Y. A. and Hannuksela, O. A. and Hasan, M. and He, H. H. and He, H. N. and He, J. Y. and He, X. Y. and He, Y. and Hernández-Cadena, S. and Hou, B. W. and Hou, C. and Hou, X. and Hu, H. B. and Hu, S. C. and Huang, C. and Huang, D. H. and Huang, J. J. and Huang, T. Q. and Huang, W. J. and Huang, X. T. and Huang, X. Y. and Huang, Y. and Huang, Y. Y. and Ji, X. L. and Jia, H. Y. and Jia, K. and Jiang, H. B. and Jiang, K. and Jiang, X. W. and Jiang, Z. J. and Jin, M. and Kaci, S. and Kang, M. M. and Karpikov, I. and Khangulyan, D. and Kuleshov, D. and Kurinov, K. and Li, B. B. and Li, Cheng and Li, Cong and Li, D. and Li, F. and Li, H. B. and Li, H. C. and Li, Jian and Li, Jie and Li, K. and Li, L. and Li, R. L. and Li, S. D. and Li, T. Y. and Li, W. L. and Li, X. R. and Li, Xin and Li, Y. and Li, Y. Z. and Li, Zhe and Li, Zhuo and Liang, E. W. and Liang, Y. F. and Lin, S. J. and Liu, B. and Liu, C. and Liu, D. and Liu, D. B. and Liu, H. and Liu, H. D. and Liu, J. and Liu, J. L. and Liu, J. R. and Liu, M. Y. and Liu, R. Y. and Liu, S. M. and Liu, W. and Liu, X. and Liu, Y. and Liu, Y. and Liu, Y. N. and Lou, Y. Q. and Luo, Q. and Luo, Y. and Lv, H. K. and Ma, B. Q. and Ma, L. L. and Ma, X. H. and Mao, J. R. and Min, Z. and Mitthumsiri, W. and Mou, G. B. and Mu, H. J. and Neronov, A. and Ng, K. C. Y. and Ni, M. Y. and Nie, L. and Ou, L. J. and Pattarakijwanich, P. and Pei, Z. Y. and Qi, J. C. and Qi, M. Y. and Qin, J. J. and Raza, A. and Ren, C. Y. and Ruffolo, D. and Sáiz, A. and Semikoz, D. and Shao, L. and Shchegolev, O. and Shen, Y. Z. and Sheng, X. D. and Shi, Z. D. and Shu, F. W. and Song, H. C. and Stenkin, Yu. V. and Stepanov, V. and Su, Y. and Sun, D. X. and Sun, H. and Sun, Q. N. and Sun, X. N. and Sun, Z. B. and Tabasam, N. H. and Takata, J. and Tam, P. H. T. and Tan, H. B. and Tang, Q. W. and Tang, R. and Tang, Z. B. and Tian, W. W. and Tong, C. N. and Wan, L. H. and Wang, C. and Wang, G. W. and Wang, H. G. and Wang, J. C. and Wang, K. and Wang, Kai and Wang, Kai and Wang, L. P. and Wang, L. Y. and Wang, L. Y. and Wang, R. and Wang, W. and Wang, X. G. and Wang, X. J. and Wang, X. Y. and Wang, Y. and Wang, Y. D. and Wang, Z. H. and Wang, Z. X. and Wang, Zheng and Wei, D. M. and Wei, J. J. and Wei, Y. J. and Wen, T. and Weng, S. S. and Wu, C. Y. and Wu, H. R. and Wu, Q. W. and Wu, S. and Wu, X. F. and Wu, Y. S. and Xi, S. Q. and Xia, J. and Xia, J. J. and Xiang, G. M. and Xiao, D. X. and Xiao, G. and Xin, Y. L. and Xing, Y. and Xiong, D. R. and Xiong, Z. and Xu, D. L. and Xu, R. F. and Xu, R. X. and Xu, W. L. and Xue, L. and Yan, D. H. and Yan, J. Z. and Yan, T. and Yang, C. W. and Yang, C. Y. and Yang, F. F. and Yang, L. L. and Yang, M. J. and Yang, R. Z. and Yang, W. X. and Yang, Z. H. and Yao, Z. G. and Ye, X. A. and Yin, L. Q. and Yin, N. and You, X. H. and You, Z. Y. and Yuan, Q. and Yue, H. and Zeng, H. D. and Zeng, T. X. and Zeng, W. and Zeng, X. T. and Zha, M. and Zhang, B. B. and Zhang, B. T. and Zhang, C. and Zhang, F. and Zhang, H. and Zhang, H. M. and Zhang, H. Y. and Zhang, J. L. and Zhang, Li and Zhang, P. F. and Zhang, P. P. and Zhang, R. and Zhang, S. R. and Zhang, S. S. and Zhang, W. Y. and Zhang, X. and Zhang, X. P. and Zhang, Yi and Zhang, Yong and Zhang, Z. P. and Zhao, J. and Zhao, L. and Zhao, L. Z. and Zhao, S. P. and Zhao, X. H. and Zhao, Z. H. and Zheng, F. and Zhong, W. J. and Zhou, B. and Zhou, H. and Zhou, J. N. and Zhou, M. and Zhou, P. and Zhou, R. and Zhou, X. X. and Zhou, X. X. and Zhu, B. Y. and Zhu, C. G. and Zhu, F. R. and Zhu, H. and Zhu, K. J. and Zou, Y. C. and Zuo, X.},
  journal = {Phys. Rev. Lett.},
  pages = {--},
  year = {2026},
  month = {Mar},
  publisher = {American Physical Society},
  doi = {10.1103/7xhp-tff7},
  url = {https://link.aps.org/doi/10.1103/7xhp-tff7}
}

@ARTICLE{Wu2018,
       author = {{Wu}, Y.~W. and {Torricelli-Ciamponi}, G. and {Massi}, M. and {Reid}, M.~J. and {Zhang}, B. and {Shao}, L. and {Zheng}, X.~W.},
        title = "{Revisiting LS I +61{\textdegree}303 with VLBI astrometry}",
      journal = {\mnras},
         year = 2018,
        month = mar,
       volume = {474},
       number = {3},
        pages = {4245-4253},
          doi = {10.1093/mnras/stx3003},
archivePrefix = {arXiv},
       eprint = {1711.07598},
 primaryClass = {astro-ph.GA},
       adsurl = {https://ui.adsabs.harvard.edu/abs/2018MNRAS.474.4245W}
}

@ARTICLE{Torres2012,
       author = {{Torres}, Diego F. and {Rea}, Nanda and {Esposito}, Paolo and {Li}, Jian and {Chen}, Yupeng and {Zhang}, Shu},
        title = "{A Magnetar-like Event from LS I +61{\textdegree}303 and Its Nature as a Gamma-Ray Binary}",
      journal = {\apj},
         year = 2012,
        month = jan,
       volume = {744},
       number = {2},
          eid = {106},
        pages = {106},
          doi = {10.1088/0004-637X/744/2/106},
archivePrefix = {arXiv},
       eprint = {1109.5008},
 primaryClass = {astro-ph.HE},
       adsurl = {https://ui.adsabs.harvard.edu/abs/2012ApJ...744..106T}
}

@ARTICLE{MaraschiTreves1981,
       author = {{Maraschi}, L. and {Treves}, A.},
        title = "{A model for LS I +61 303.}",
      journal = {\mnras},
         year = 1981,
        month = jan,
       volume = {194},
        pages = {1P-5},
          doi = {10.1093/mnras/194.1.1P},
       adsurl = {https://ui.adsabs.harvard.edu/abs/1981MNRAS.194P...1M}
}

@article{Acciari2008,
doi = {10.1086/587736},
url = {https://doi.org/10.1086/587736},
year = {2008},
month = {jun},
publisher = {},
volume = {679},
number = {2},
pages = {1427},
author = {Acciari, V. A. and Beilicke, M. and Blaylock, G. and Bradbury, S. M. and Buckley, J. H. and Bugaev, V. and Butt, Y. and Byrum, K. L. and Celik, O. and Cesarini, A. and Ciupik, L. and Chow, Y. C. K. and Cogan, P. and Colin, P. and Cui, W. and Daniel, M. K. and Duke, C. and Ergin, T. and Falcone, A. D. and Fegan, S. J. and Finley, J. P. and Fortin, P. and Fortson, L. F. and Gall, D. and Gibbs, K. and Gillanders, G. H. and Grube, J. and Guenette, R. and Hanna, D. and Hays, E. and Holder, J. and Horan, D. and Hughes, S. B. and Hui, C. M. and Humensky, T. B. and Kaaret, P. and Kieda, D. B. and Kildea, J. and Konopelko, A. and Krawczynski, H. and Krennrich, F. and Lang, M. J. and LeBohec, S. and Lee, K. and Maier, G. and McCann, A. and McCutcheon, M. and Millis, J. and Moriarty, P. and Mukherjee, R. and Nagai, T. and Ong, R. A. and Pandel, D. and Perkins, J. S. and Pizlo, F. and Pohl, M. and Quinn, J. and Ragan, K. and Reynolds, P. T. and Rose, H. J. and Schroedter, M. and Sembroski, G. H. and Smith, A. W. and Steele, D. and Swordy, S. P. and Toner, J. A. and Valcarcel, L. and Vassiliev, V. V. and Wagner, R. and Wakely, S. P. and Ward, J. E. and Weekes, T. C. and Weinstein, A. and White, R. J. and Williams, D. A. and Wissel, S. A. and Wood, M. and Zitzer, B.},
title = {VERITAS Observations of the γ-Ray Binary LS I +61 303},
journal = {The Astrophysical Journal}
}

@ARTICLE{Miralles2023,
       author = {{L{\'o}pez-Miralles}, J. and {Motta}, Sara E. and {Migliari}, S. and {Jaron}, F.},
        title = "{Rapid X-ray variability of the gamma-ray binary LS I +61{\textdegree}303}",
      journal = {\mnras},
         year = 2023,
        month = aug,
       volume = {523},
       number = {3},
        pages = {4282-4293},
          doi = {10.1093/mnras/stad1658},
archivePrefix = {arXiv},
       eprint = {2305.18580},
 primaryClass = {astro-ph.HE},
       adsurl = {https://ui.adsabs.harvard.edu/abs/2023MNRAS.523.4282L}
}

@ARTICLE{Grundstrom2007,
       author = {{Grundstrom}, E.~D. and {Caballero-Nieves}, S.~M. and {Gies}, D.~R. and {Huang}, W. and {McSwain}, M.~V. and {Rafter}, S.~E. and {Riddle}, R.~L. and {Williams}, S.~J. and {Wingert}, D.~W.},
        title = "{Joint H{\ensuremath{\alpha}} and X-Ray Observations of Massive X-Ray Binaries. II. The Be X-Ray Binary and Microquasar LS I +61 303}",
      journal = {\apj},
         year = 2007,
        month = feb,
       volume = {656},
       number = {1},
        pages = {437-443},
          doi = {10.1086/510509},
archivePrefix = {arXiv},
       eprint = {astro-ph/0610608},
 primaryClass = {astro-ph},
       adsurl = {https://ui.adsabs.harvard.edu/abs/2007ApJ...656..437G}
}

@ARTICLE{Jaron2024,
       author = {{Jaron}, F. and {Kiehlmann}, S. and {Readhead}, A.~C.~S.},
        title = "{Owens Valley Radio Observatory monitoring of LS I +61{\textdegree}303 completes three cycles of the super-orbital modulation}",
      journal = {\aap},
         year = 2024,
        month = mar,
       volume = {683},
          eid = {A228},
        pages = {A228},
          doi = {10.1051/0004-6361/202347871},
archivePrefix = {arXiv},
       eprint = {2402.07719},
 primaryClass = {astro-ph.HE},
       adsurl = {https://ui.adsabs.harvard.edu/abs/2024A&A...683A.228J}
}

@ARTICLE{Li2011,
       author = {{Li}, Jian and {Torres}, Diego F. and {Zhang}, Shu and {Chen}, Yupeng and {Hadasch}, Daniela and {Ray}, Paul S. and {Kretschmar}, Peter and {Rea}, Nanda and {Wang}, Jianmin},
        title = "{Long-term X-Ray Monitoring of LS I +61{\textdegree}303: Analysis of Spectral Variability and Flares}",
      journal = {\apj},
         year = 2011,
        month = jun,
       volume = {733},
       number = {2},
          eid = {89},
        pages = {89},
          doi = {10.1088/0004-637X/733/2/89},
archivePrefix = {arXiv},
       eprint = {1103.4205},
 primaryClass = {astro-ph.HE},
       adsurl = {https://ui.adsabs.harvard.edu/abs/2011ApJ...733...89L}
}

@ARTICLE{Smith2009,
       author = {{Smith}, A. and {Kaaret}, P. and {Holder}, J. and {Falcone}, A. and {Maier}, G. and {Pandel}, D. and {Stroh}, M.},
        title = "{Long-Term X-Ray Monitoring of the TeV Binary LS I +61 303 With the Rossi X-Ray Timing Explorer}",
      journal = {\apj},
         year = 2009,
        month = mar,
       volume = {693},
       number = {2},
        pages = {1621-1627},
          doi = {10.1088/0004-637X/693/2/1621},
archivePrefix = {arXiv},
       eprint = {0809.4254},
 primaryClass = {astro-ph},
       adsurl = {https://ui.adsabs.harvard.edu/abs/2009ApJ...693.1621S}
}

@ARTICLE{Archambault2016,
doi = {10.3847/2041-8205/817/1/L7},
url = {https://doi.org/10.3847/2041-8205/817/1/L7},
year = {2016},
month = {jan},
publisher = {The American Astronomical Society},
volume = {817},
number = {1},
pages = {L7},
author = {Archambault, S. and Archer, A. and Aune, T. and Barnacka, A. and Benbow, W. and Bird, R. and Buchovecky, M. and Buckley, J. H. and Bugaev, V. and Byrum, K. and Cardenzana, J. V and Cerruti, M. and Chen, X. and Ciupik, L. and Collins-Hughes, E. and Connolly, M. P. and Cui, W. and Dickinson, H. J. and Dumm, J. and Eisch, J. D. and Falcone, A. and Feng, Q. and Finley, J. P. and Fleischhack, H. and Flinders, A. and Fortin, P. and Fortson, L. and Furniss, A. and Gillanders, G. H. and Griffin, S. and Grube, J. and Gyuk, G. and Hütten, M. and Håkansson, N. and Hanna, D. and Holder, J. and Humensky, T. B. and Johnson, C. A. and Kaaret, P. and Kar, P. and Kelley-Hoskins, N. and Kertzman, M. and Khassen, Y. and Kieda, D. and Krause, M. and Krennrich, F. and Kumar, S. and Lang, M. J. and Maier, G. and McArthur, S. and McCann, A. and Meagher, K. and Millis, J. and Moriarty, P. and Mukherjee, R. and Nieto, D. and O’Brien, S. and Bhróithe, A. O’Faoláin de and Ong, R. A. and Otte, A. N. and Pandel, D. and Park, N. and Pelassa, V. and Pohl, M. and Popkow, A. and Pueschel, E. and Quinn, J. and Ragan, K. and Reynolds, P. T. and Richards, G. T. and Roache, E. and Rousselle, J. and Rulten, C. and Santander, M. and Sembroski, G. H. and Shahinyan, K. and Smith, A. W. and Staszak, D. and Telezhinsky, I. and Tucci, J. V. and Tyler, J. and Vincent, S. and Wakely, S. P. and Weiner, O. M. and Weinstein, A. and Wilhelm, A. and Williams, D. A. and Zitzer, B.},
title = {EXCEPTIONALLY BRIGHT TEV FLARES FROM THE BINARY LS I +61° 303},
journal = {The Astrophysical Journal Letters}
}

@ARTICLE{Albert2006,
       author = {{Albert}, J. and {Aliu}, E. and {Anderhub}, H. and {Antoranz}, P. and {Armada}, A. and {Asensio}, M. and {Baixeras}, C. and {Barrio}, J.~A. and {Bartelt}, M. and {Bartko}, H. and {Bastieri}, D. and {Bavikadi}, S.~R. and {Bednarek}, W. and {Berger}, K. and {Bigongiari}, C. and {Biland}, A. and {Bisesi}, E. and {Bock}, R.~K. and {Bordas}, P. and {Bosch-Ramon}, V. and {Bretz}, T. and {Britvitch}, I. and {Camara}, M. and {Carmona}, E. and {Chilingarian}, A. and {Ciprini}, S. and {Coarasa}, J.~A. and {Commichau}, S. and {Contreras}, J.~L. and {Cortina}, J. and {Curtef}, V. and {Danielyan}, V. and {Dazzi}, F. and {De Angelis}, A. and {de los Reyes}, R. and {De Lotto}, B. and {Domingo-Santamar{\'\i}a}, E. and {Dorner}, D. and {Doro}, M. and {Errando}, M. and {Fagiolini}, M. and {Ferenc}, D. and {Fern{\'a}ndez}, E. and {Firpo}, R. and {Flix}, J. and {Fonseca}, M.~V. and {Font}, L. and {Fuchs}, M. and {Galante}, N. and {Garczarczyk}, M. and {Gaug}, M. and {Giller}, M. and {Goebel}, F. and {Hakobyan}, D. and {Hayashida}, M. and {Hengstebeck}, T. and {H{\"o}hne}, D. and {Hose}, J. and {Hsu}, C.~C. and {Isar}, P.~G. and {Jacon}, P. and {Kalekin}, O. and {Kosyra}, R. and {Kranich}, D. and {Laatiaoui}, M. and {Laille}, A. and {Lenisa}, T. and {Liebing}, P. and {Lindfors}, E. and {Lombardi}, S. and {Longo}, F. and {L{\'o}pez}, J. and {L{\'o}pez}, M. and {Lorenz}, E. and {Lucarelli}, F. and {Majumdar}, P. and {Maneva}, G. and {Mannheim}, K. and {Mansutti}, O. and {Mariotti}, M. and {Mart{\'\i}nez}, M. and {Mase}, K. and {Mazin}, D. and {Merck}, C. and {Meucci}, M. and {Meyer}, M. and {Miranda}, J.~M. and {Mirzoyan}, R. and {Mizobuchi}, S. and {Moralejo}, A. and {Nilsson}, K. and {O{\~n}a-Wilhelmi}, E. and {Ordu{\~n}a}, R. and {Otte}, N. and {Oya}, I. and {Paneque}, D. and {Paoletti}, R. and {Paredes}, J.~M. and {Pasanen}, M. and {Pascoli}, D. and {Pauss}, F. and {Pavel}, N. and {Pegna}, R. and {Persic}, M. and {Peruzzo}, L. and {Piccioli}, A. and {Poller}, M. and {Pooley}, G. and {Prandini}, E. and {Raymers}, A. and {Rhode}, W. and {Rib{\'o}}, M. and {Rico}, J. and {Riegel}, B. and {Rissi}, M. and {Robert}, A. and {Romero}, G.~E. and {R{\"u}gamer}, S. and {Saggion}, A. and {S{\'a}nchez}, A. and {Sartori}, P. and {Scalzotto}, V. and {Scapin}, V. and {Schmitt}, R. and {Schweizer}, T. and {Shayduk}, M. and {Shinozaki}, K. and {Shore}, S.~N. and {Sidro}, N. and {Sillanp{\"a}{\"a}}, A. and {Sobczynska}, D. and {Stamerra}, A. and {Stark}, L.~S. and {Takalo}, L. and {Temnikov}, P. and {Tescaro}, D. and {Teshima}, M. and {Tonello}, N. and {Torres}, A. and {Torres}, D.~F. and {Turini}, N. and {Vankov}, H. and {Vitale}, V. and {Wagner}, R.~M. and {Wibig}, T. and {Wittek}, W. and {Zanin}, R. and {Zapatero}, J.},
        title = "{Variable Very-High-Energy Gamma-Ray Emission from the Microquasar LS I +61 303}",
      journal = {Science},
         year = 2006,
        month = jun,
       volume = {312},
       number = {5781},
        pages = {1771-1773},
          doi = {10.1126/science.1128177},
archivePrefix = {arXiv},
       eprint = {astro-ph/0605549},
 primaryClass = {astro-ph},
       adsurl = {https://ui.adsabs.harvard.edu/abs/2006Sci...312.1771A}
}

@ARTICLE{Dubus2013,
       author = {{Dubus}, G.},
        title = "{Gamma-ray binaries and related systems}",
      journal = {\aapr},
         year = 2013,
        month = aug,
       volume = {21},
          eid = {64},
        pages = {64},
          doi = {10.1007/s00159-013-0064-5},
archivePrefix = {arXiv},
       eprint = {1307.7083},
 primaryClass = {astro-ph.HE},
       adsurl = {https://ui.adsabs.harvard.edu/abs/2013A&ARv..21...64D}
}

@ARTICLE{Lindegren2021,
       author = {{Lindegren}, L. and {Klioner}, S.~A. and {Hern{\'a}ndez}, J. and {Bombrun}, A. and {Ramos-Lerate}, M. and {Steidelm{\"u}ller}, H. and {Bastian}, U. and {Biermann}, M. and {de Torres}, A. and {Gerlach}, E. and {Geyer}, R. and {Hilger}, T. and {Hobbs}, D. and {Lammers}, U. and {McMillan}, P.~J. and {Stephenson}, C.~A. and {Casta{\~n}eda}, J. and {Davidson}, M. and {Fabricius}, C. and {Gracia-Abril}, G. and {Portell}, J. and {Rowell}, N. and {Teyssier}, D. and {Torra}, F. and {Bartolom{\'e}}, S. and {Clotet}, M. and {Garralda}, N. and {Gonz{\'a}lez-Vidal}, J.~J. and {Torra}, J. and {Abbas}, U. and {Altmann}, M. and {Anglada Varela}, E. and {Balaguer-N{\'u}{\~n}ez}, L. and {Balog}, Z. and {Barache}, C. and {Becciani}, U. and {Bernet}, M. and {Bertone}, S. and {Bianchi}, L. and {Bouquillon}, S. and {Brown}, A.~G.~A. and {Bucciarelli}, B. and {Busonero}, D. and {Butkevich}, A.~G. and {Buzzi}, R. and {Cancelliere}, R. and {Carlucci}, T. and {Charlot}, P. and {Cioni}, M.-R.~L. and {Crosta}, M. and {Crowley}, C. and {del Peloso}, E.~F. and {del Pozo}, E. and {Drimmel}, R. and {Esquej}, P. and {Fienga}, A. and {Fraile}, E. and {Gai}, M. and {Garcia-Reinaldos}, M. and {Guerra}, R. and {Hambly}, N.~C. and {Hauser}, M. and {Jan{\ss}en}, K. and {Jordan}, S. and {Kostrzewa-Rutkowska}, Z. and {Lattanzi}, M.~G. and {Liao}, S. and {Licata}, E. and {Lister}, T.~A. and {L{\"o}ffler}, W. and {Marchant}, J.~M. and {Masip}, A. and {Mignard}, F. and {Mints}, A. and {Molina}, D. and {Mora}, A. and {Morbidelli}, R. and {Murphy}, C.~P. and {Pagani}, C. and {Panuzzo}, P. and {Pe{\~n}alosa Esteller}, X. and {Poggio}, E. and {Re Fiorentin}, P. and {Riva}, A. and {Sagrist{\`a} Sell{\'e}s}, A. and {Sanchez Gimenez}, V. and {Sarasso}, M. and {Sciacca}, E. and {Siddiqui}, H.~I. and {Smart}, R.~L. and {Souami}, D. and {Spagna}, A. and {Steele}, I.~A. and {Taris}, F. and {Utrilla}, E. and {van Reeven}, W. and {Vecchiato}, A.},
        title = "{Gaia Early Data Release 3. The astrometric solution}",
      journal = {\aap},
         year = 2021,
        month = may,
       volume = {649},
          eid = {A2},
        pages = {A2},
          doi = {10.1051/0004-6361/202039709},
archivePrefix = {arXiv},
       eprint = {2012.03380},
 primaryClass = {astro-ph.IM},
       adsurl = {https://ui.adsabs.harvard.edu/abs/2021A&A...649A...2L}
}

@ARTICLE{Tavani1997,
       author = {{Tavani}, M. and {Arons}, J.},
        title = "{Theory of High-Energy Emission from the Pulsar/Be Star System PSR 1259-63. I. Radiation Mechanisms and Interaction Geometry}",
      journal = {\apj},
         year = 1997,
        month = mar,
       volume = {477},
       number = {1},
        pages = {439-464},
          doi = {10.1086/303676},
archivePrefix = {arXiv},
       eprint = {astro-ph/9609086},
 primaryClass = {astro-ph},
       adsurl = {https://ui.adsabs.harvard.edu/abs/1997ApJ...477..439T}
}

@ARTICLE{Johnston1994,
       author = {{Johnston}, S. and {Manchester}, R.~N. and {Lyne}, A.~G. and {Nicastro}, L. and {Spyromilio}, J.},
        title = "{Radio and Optical Observations of the PSR:B1259-63 / SS:2883 Be-Star Binary System}",
      journal = {\mnras},
         year = 1994,
        month = may,
       volume = {268},
        pages = {430},
          doi = {10.1093/mnras/268.2.430},
       adsurl = {https://ui.adsabs.harvard.edu/abs/1994MNRAS.268..430J}
}

@ARTICLE{Xingxing2021,
       author = {{Xingxing}, Hu and {Jumpei}, Takata},
        title = "{Polarization Study of Gamma-ray Binary Systems}",
      journal = {\apj},
         year = 2021,
        month = dec,
       volume = {922},
       number = {2},
          eid = {260},
        pages = {260},
          doi = {10.3847/1538-4357/ac273b},
archivePrefix = {arXiv},
       eprint = {2111.04300},
 primaryClass = {astro-ph.HE},
       adsurl = {https://ui.adsabs.harvard.edu/abs/2021ApJ...922..260X}
}

@ARTICLE{Ramsey2022,
       author = {{Ramsey}, Brian D. and {Bongiorno}, Stephen D. and {Kolodziejczak}, Jeffery J. and {Kilaru}, Kiranmayee and {Alexander}, Cheryl and {Baumgartner}, Wayne H. and {Breeding}, Shawn and {Elsner}, Ronald F. and {Le Roy}, Shelley and {McCracken}, Jeff and {Mitsuishi}, Ikuyuki and {O'Dell}, Stephen L. and {Pavelitz}, Steven D. and {Ranganathan}, Jaganathan and {Sanchez}, Javier and {Speegle}, Chet O. and {Thomas}, Nicholas and {Weddendorf}, Bruce and {Weisskopf}, Martin C.},
        title = "{Optics for the imaging x-ray polarimetry explorer}",
      journal = {Journal of Astronomical Telescopes, Instruments, and Systems},
         year = 2022,
        month = apr,
       volume = {8},
          eid = {024003},
        pages = {024003},
          doi = {10.1117/1.JATIS.8.2.024003},
       adsurl = {https://ui.adsabs.harvard.edu/abs/2022JATIS...8b4003R}
}

@ARTICLE{Soffitta2021,
       author = {{Soffitta}, Paolo and {Baldini}, Luca and {Bellazzini}, Ronaldo and {Costa}, Enrico and {Latronico}, Luca and {Muleri}, Fabio and {Del Monte}, Ettore and {Fabiani}, Sergio and {Minuti}, Massimo and {Pinchera}, Michele and {Sgro'}, Carmelo and {Spandre}, Gloria and {Trois}, Alessio and {Amici}, Fabrizio and {Andersson}, Hans and {Attina'}, Primo and {Bachetti}, Matteo and {Barbanera}, Mattia and {Borotto}, Fabio and {Brez}, Alessandro and {Brienza}, Daniele and {Caporale}, Ciro and {Cardelli}, Claudia and {Carpentiero}, Rita and {Castellano}, Simone and {Castronuovo}, Marco and {Cavalli}, Luca and {Cavazzuti}, Elisabetta and {Ceccanti}, Marco and {Centrone}, Mauro and {Ciprini}, Stefano and {Citraro}, Saverio and {D'Amico}, Fabio and {D'Alba}, Elisa and {Di Cosimo}, Sergio and {Di Lalla}, Niccolo' and {Di Marco}, Alessandro and {Di Persio}, Giuseppe and {Donnarumma}, Immacolata and {Evangelista}, Yuri and {Ferrazzoli}, Riccardo and {Hayato}, Asami and {Kitaguchi}, Takao and {La Monaca}, Fabio and {Lefevre}, Carlo and {Loffredo}, Pasqualino and {Lorenzi}, Paolo and {Lucchesi}, Leonardo and {Magazzu}, Carlo and {Maldera}, Simone and {Manfreda}, Alberto and {Mangraviti}, Elio and {Marengo}, Marco and {Matt}, Giorgio and {Mereu}, Paolo and {Morbidini}, Alfredo and {Mosti}, Federico and {Nakano}, Toshio and {Nasimi}, Hikmat and {Negri}, Barbara and {Nenonen}, Seppo and {Nuti}, Alessio and {Orsini}, Leonardo and {Perri}, Matteo and {Pesce-Rollins}, Melissa and {Piazzolla}, Raffaele and {Pilia}, Maura and {Profeti}, Alessandro and {Puccetti}, Simonetta and {Rankin}, John and {Ratheesh}, Ajay and {Rubini}, Alda and {Santoli}, Francesco and {Sarra}, Paolo and {Scalise}, Emanuele and {Sciortino}, Andrea and {Tamagawa}, Toru and {Tardiola}, Marcello and {Tobia}, Antonino and {Vimercati}, Marco and {Xie}, Fei},
        title = "{The Instrument of the Imaging X-Ray Polarimetry Explorer}",
      journal = {\aj},
         year = 2021,
        month = nov,
       volume = {162},
       number = {5},
          eid = {208},
        pages = {208},
          doi = {10.3847/1538-3881/ac19b0},
archivePrefix = {arXiv},
       eprint = {2108.00284},
 primaryClass = {astro-ph.IM},
       adsurl = {https://ui.adsabs.harvard.edu/abs/2021AJ....162..208S}
}
\bibliographystyle{aasjournal}


\end{document}